# Interfacial Temperature and Density Discontinuities for Phase-Change Heat Transfer With Non-condensable Gas

Gang Chen

Department of Mechanical Engineering
Massachusetts Institute of Technology
Cambridge, MA 02139

**Abstract**

In recent prior work, the author derived interfacial mass and heat flux conditions for phase-change processes. The mass flux condition is identical to the Schrage equation, but the additional heat flux expression enables one to couple the interface to the continua in both the liquid and the vapor phases and compute the interfacial temperature and density discontinuities. However, questions exist on how to treat phase change heat transfer in the presence of non-condensable gases. In this work, the author shows that the same set of interfacial conditions can be used to account for the presence of non-condensable gases. Although the mass flux of non-condensable gas is zero, their presence impacts the heat transfer. For evaporation, when the presence of the non-condensable gas is small, temperature and density discontinuities persist across the interface, as well as inverted temperature distributions. For condensation, however, no temperature inversion happens in the presence of a small amount of non-condensable gas and the interfacial temperature jump is significantly smaller. When a large amount of non-condensable gas is present, such as for evaporation into and condensation from air, the temperature discontinuities at the interface are significantly smaller and no temperature inversion happens. For evaporation driven purely by humidity difference, temperature inversion and discontinuity still exist. Results from this work will benefit the modeling of phase change processes in the presence of non-condensable gases, evaporative cooling in air, air-gap distillation, atmospheric water harvesting, and other applications.

I. Introduction

Phase-change heat transfer remains a challenging topic despite progress made since Nusselt's pioneering work [1–6], which is further complicated by presence of non-condensable gas [7–11]. Applications such as cloud formation, solar interfacial-evaporation and desalination [12,13], atmospheric water harvesting [14], drying [15] call for the inclusion of the impact of non-condensable gases. Even in closed systems intended to be for pure vapor operation such as heat pipes, vapor chambers, and condensers in power plants [10,11], non-condensable gases accumulate due to degassing, which significantly impacts the device and system performance. It is well-known that the presence of non-condensable gas significantly impedes condensation heat transfer, but less so evaporation heat transfer. Past studies on non-condensable gas effects are extensive, including kinetic theory approaches based on solving the Boltzmann transport equation (BTE) [16–25], the lattice Boltzmann method [26], Monte Carlo and molecular dynamics simulations [27,28], the Navier-Stokes equations-based continuum approaches [10,11,29–31], and heat-and-mass transfer analogy [32]. Most of the engineering treatments assumed interfacial temperature continuity, despite the fact that temperature discontinuities had been demonstrated experimentally in the evaporation and condensation of pure water [33–36]. Questions remain how the presence of non-condensable gas impact such the interfacial discontinuities.

In limited treatments of phase change process including the interface temperature discontinuities, the Schrage equation for mass flux is often used [2,37].

$$\dot{m} = \frac{2\alpha}{2-\alpha}\sqrt{\frac{R}{2\pi M}}\left[\rho_s(T_s)\sqrt{T_s} - \rho_v\sqrt{T_v}\right] \qquad (1)$$

where $\alpha$ is the accommodation coefficient, $\rho$ the density, R the universal ideal gas constant, M the molecular weight, and T the temperature. The subscript "s" represents the properties of the saturated vapor phase on the liquid surface, and "v" the vapor phase properties at the outer edge of the Knudsen layer, which is of the order of a few mean free path lengths [38,39]. This thickness is neglected and hence "v" can be considered properties of the vapor phase immediately outside the liquid surface.

To apply the Schrage equation in practical cases, one needs additional conditions so that $T_s$, $T_v$, $\rho_v$ can be uniquely determined. In the past, this problem was often solved by assuming either $T_s=T_v$ or $\rho_s(T_s) = \rho_v$, which cannot explain the experimentally measured interfacial discontinuities [40]. The author recently addressed this problem by deriving a heat flux expression at the interface starting from kinetic theory [41,42],

$$q = \frac{4\alpha}{2-\alpha}\frac{R}{M}\sqrt{\frac{R}{2\pi M}}\left[\rho_s(T_s)T_s^{3/2} - \rho_v T_v^{3/2}\right] \qquad (2)$$

The above equation can be combined with Eq. (1) to couple the continua descriptions in both the liquid and the vapor phases, allowing one to determine temperature and density (as well as pressure) discontinuities at interfaces during phase change heat transfer. The author had shown that the solutions obtained for evaporation and condensation of a pure substance at an interface can reasonably explain past experiments that had defied modelling efforts before [41,42]. For the two parallel-plate problem of evaporation on one side and condensation on the other, these interface conditions coupled to the continuum treatments of the liquid and vapor phases also lead to the classical result of inverted temperature profile predicted from the kinetic theory, i.e., colder vapor temperature at the evaporating interface than at the condensing interface [43].

The interfacial conditions, Eqs. (1) and (2), however, were established for pure substances only. In theory, one can follow the same strategies used in deriving these conditions, i.e., starting from the BTE under the BGK approximation [24,38,44,45], to derive similar interfacial conditions. However, extension of the BGK approximation to multicomponent gas mixtures is not straightforward and has been a subject of continued investigations [16–24,46,47]. There are two distinct schools of thought [21,48]. One school splits the scattering of molecules into intra-species and inter-species, each with a corresponding BGK-type of expression with its own scattering rates, velocity, and temperature [18,23,24,49]. We will call this the "two-term approach." The other school groups intra and inter molecular scattering into one BGK-type of relaxation form using a common drift velocity [16,17,19,20,50], which we will call the "one-term approach." Direct solution of coupled integral BTE has also been explored [47,51]. While most of these references study the methodology, the work of Aoki et al. [47] suggests that even the presence of small amount of non-condensable gas has significant impacts on heat transfer for the two parallel plate problem. These studies, including using molecular dynamics simulations, have shown that the Schrage equation generally holds well even with non-condensable gas [52–55]. However, the Schrage equation alone cannot determine the interfacial temperature discontinuity. We need to develop a parallel treatment for the interfacial heat flux.

In this manuscript, the author will extend the interface conditions derived for the phase change heat transfer of a pure substance to gas mixtures and use the extended interface conditions to study how the presence of non-condensable gas impacts phase change heat transfer. The extended boundary conditions allow coupling transport at interfaces to the transport in the bulk regions described by the continuum equations. Using one-dimensional condensation and evaporation problems as an example, the paper will examine the influence of the presence of the non-condensable gas on phase change heat transfer, especially on the temperature discontinuity and inversion phenomena that the author discussed for evaporation and condensation in pure phase [41–43]. For evaporation, the influence due to the presence of the non-condensable gas is small (a quantitative discussion on how small is small will be presented later), temperature and density discontinuities persist across the interface, and temperature distributions are inverted, i.e., the gas phase temperature increases with increasing distance from the interface. Such inverted temperature distributions mean that heat is conducted back to the interface while convection takes heat away from the interface in the gaseous phase. However, the presence of a small amount of non-condensable gas significantly changes the temperature profile during condensation: the temperature distributions are no longer inverted and the interfacial temperature jump diminishes. For situations such as evaporation into and condensation from air, i.e., when the non-condensable gas fraction is large, the temperature discontinuities at the interface are significantly smaller and no temperature inversion happens, while for evaporation driven by humidity difference only, a temperature inversion still exists. The boundary conditions arrived in this work are applicable for both evaporation and condensation, enabling better modeling for applications such as evaporative cooling and atmospheric water harvesting, air-gap distillation and desalination, cloud formation, solar-interfacial evaporation, and for evaluating the impact of the presence of minute amount of non-condensable gas for phase change heat transfer.

## 2. Interface Conditions Including Non-condensable Gas

We choose the one-term approach for the BGK approximation of the BTE [16,17,19,20,49,50] as this approach was shown to be consistent with Fick's law and other constraints required for the BTE. Since the interface conditions we aim to develop basically neglect the details of the transport in the Knudsen layer, the choice of either the one-term or the two-term approaches may not be important for the final results. The simplicity of the one term approach is another reason behind our choice.

The idea in the one-term approach is that one "global" operator in the BGK approximation is used for each species i, taking into account its collision with all the species including itself [19]. We will follow the work of Brull et al. [17], whose model was an extension of Andries et al. [19], since the former has shown consistency with the Navier-Stokes equation, Fick's law and other transport coefficients, and the Onsager relations. We consider transport to be along the z-direction as shown in Fig. 1 for simplicity, although the relations we derive should be generally applicable, including curved surfaces and multidimensional transport. The BGK-type of the BTE for each species is

$$v_{zi}\frac{\partial f_i}{\partial z} = -\frac{f_i - f_{oi}}{\tau_i} \qquad (3)$$

with $f_{oi}$ the displaced Maxwell-Boltzmann distribution given by

$$f_{oi} = n_i \left[\frac{M_i}{2\pi RT}\right]^{3/2} exp\left\{-\frac{M_i(v-u)^2}{2RT}\right\} \tag{4}$$

where $n_i$ is the molar density of molecules, $m_i$ the mass, $\tau_i$ the relaxation time, and the subscript represents the i-th species; **v** is the random molecular velocity and **u** is the average drift velocity, which is the same for all species. For the geometry we consider, $\boldsymbol{u} = u(z)\hat{z}$. Note that in the original work of Andries et al. [19], the temperature of each species was assumed to be different. Brull et al. [17] instead used same T* for all $f_{oi}$, but pointed out that T* is different from the thermodynamic temperature T, by an order $m_i(\boldsymbol{u_i}-\boldsymbol{u})^2/k_B$, which should be only a small correction as long as the drift velocities of the molecules are small, i.e., low Mach number flows, and hence we will simply use T to represent the temperature.

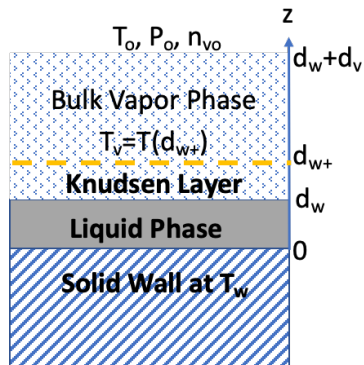

Figure 1: Modelled configuration. Liquid layer thickness is $d_w$, $d_{w+}$ is at the edge of the Knudsen layer, which is approximated as zero thickness.

We consider a two species system, with i=v representing the vapor phase and i=a the non-condensable phase such as air. Following same procedure as detailed in the appendix of Ref. [41], we arrive at the following expressions for molar mass flux and heat flux at interface for each species:

$$\dot{N}_i = \frac{2\alpha_i}{2-\alpha_i}\sqrt{\frac{R}{2\pi M_i}}\left[n_{is}\sqrt{T_s} - n_{i1}\sqrt{T_1}\right] \tag{5}$$

$$q_i = \frac{4\alpha_i}{2-\alpha_i}R\sqrt{\frac{R}{2\pi M_i}}\left[n_{is}T_s^{\frac{3}{2}} - n_{i1}T_1^{\frac{3}{2}}\right] \tag{6}$$

where $\dot{N}_i$ is the molar flux of the i-th species, $T_s$ is the liquid temperature at the interface and $n_{is}$ the corresponding molar density of the i-th species on the liquid surface, $n_{i1}$ and $T_1$ are respectively the gas molar density and temperature at the outer edge of the Knudsen layer ($d_{w+}$ in Fig.1), which is treated as zero thickness. Note we have resorted to molar flux rather than mass flux, and the reason for this choice will be explained later. For i=v, $n_{vs}$ is determined by the interface temperature $T_s$, corresponding to the saturation density $\rho_{vs}(T_s)$. For i=a, the molar density of the non-condensable gas on the surface $n_{as}$ is not known. However, we can use the condition $\dot{N}_a = 0$ to relate it to $n_{a1}$,

$$n_{as}\sqrt{T_s} = n_{a1}\sqrt{T_1} \tag{7}$$

The net interfacial mass and heat fluxes for a two species system are thus

$$\dot{N} = \dot{N}_v = \frac{2\alpha_v}{2-\alpha_v}\sqrt{\frac{R}{2\pi M_v}}\left[n_{vs}\sqrt{T_s} - n_{v1}\sqrt{T_1}\right] \tag{8}$$

$$q = q_v + q_a = 2RT_1\dot{N}_v + \left\{\frac{4\alpha_v}{2-\alpha_v}R\sqrt{\frac{R}{2\pi M_v}}n_{vs} + \frac{4\alpha_a}{2-\alpha_a}R\sqrt{\frac{R}{2\pi M_a}}n_{as}\right\}\sqrt{T_s}(T_s - T_1) \tag{9}$$

Equations (7)-(9) are the main results of this paper. These boundary conditions are sufficient for us to determine interfacial discontinuities in temperature as well as the vapor and gas densities.

## 3. Coupling to Continuums on Both Sides of Interface

With the above interfacial conditions, we can couple the continuum descriptions for transport on both sides of the interface. We consider steady-state evaporation or condensation above a horizontal surface as shown in Fig. 1 for simplicity, although the relations we derive will be generally for other configurations. For the liquid side, we can use either pure heat conduction, or an effective convective heat transfer coefficient to represent heat supplied from the liquid side to the interface. For 1D heat conduction with fixed wall temperature, heat conduction to the interface is

$$q_w = k_w \frac{T_w - T_s}{d} = h_w(T_w - T_s) \tag{10}$$

where the second expression can be used for convective heat transfer with the appropriate convective heat transfer coefficient in place of $h_w = k_w/d$.

For the vapor side, Eqs. (3) and (4) lead to different average velocities for the two species. Using these equations, and following same procedures as outlined in Ref. [41], we arrive at the molar flux for species $v$ as

$$\dot{N}_v = n_v \boldsymbol{u} - \tau_v \frac{R}{M_v} T n_v \frac{d\ln(n_v T)}{dz} = n_v \boldsymbol{u} - \tau_v \frac{R}{M_v} nT \frac{d\chi_v}{dz} = n_v \boldsymbol{u} - nD_{va} \frac{d\chi_v}{dz} \tag{11}$$

where n the average molar concentration of the mixture $n = n_v + n_a$, $\chi_v = n_v/n$ is the mole fraction, and $D_{va}$ the diffusivity of species v in a. In the second step of Eq. (11), we used the ideal gas relation $nRT = P$, where P is the pressure. The pressure is uniform since we are ignoring the momentum equation, as is typically done for mass transfer [32]. The first term in Eq. (11) is convective mass flux and the second term the diffusion flux for the vapor species. Similarly, for the non-condensable species a, we have

$$\dot{N}_a = n_a u - nD_{av} \frac{d\chi_a}{dz} \tag{12}$$

where $D_{va}=D_{av}=D= \tau_v \frac{RT}{M_v}$, which also means $\tau_v/M_v = \tau_a/M_a$. One can check that the diffusion fluxes $J_v = -nD \frac{d\chi_v}{dz}$ and $J_a = -nD \frac{d\chi_a}{dz}$ satisfy the condition $J_v + J_a = 0$, which is a requirement of the continuity equation,

$$\nabla \cdot (n\boldsymbol{u}) = 0 \tag{13}$$

With the solution

$$\dot{N} = nu = constant \tag{14}$$

Note although most continuity equation is based on mass density rather than molar concentration, the author finds that it is easier to use molar concentration. This is consistent with treatment in the Brull et al. [17]. This choice will become clearer later.

Following same procedure as in Ref. [41], one can show that the heat fluxes carried by each species are

$$q_v = \frac{5}{2}RTn_v u - \frac{5}{2}RTnD\frac{d\chi_v}{dz} - \frac{5}{2}Rn_v D\frac{dT}{dz} \tag{15}$$

$$q_a = \frac{5}{2}RTn_a u - \frac{5}{2}RTnD\frac{d\chi_a}{dz} - \frac{5}{2}Rn_a D\frac{dT}{dz} \tag{16}$$

where pressure is again assumed to be constant. The total heat flux is then the sum of the two

$$q = \frac{5}{2}RTnu - \frac{5}{2}RnD\frac{dT}{dz} = c_p MT\dot{N} - k\frac{dT}{dz} = C_p T\dot{N} - k\frac{dT}{dz} \tag{17}$$

where k the thermal conductivity k=$\frac{5}{2}RnD$, M the average molar weight $M = M_v \chi_v + M_a \chi_a$. The above equation leveraged the fact $\chi_v + \chi_a$ =1. The second step in Eq.(17) replaced $\frac{5}{2}R$ with the constant pressure specific heat $c_p$ and the third step used $C_p = c_p M$ to represent molar specific heat. The derivation process shows that $C_p$ is a constant. If one had used mass-based expression, it is easy to create the impression that $c_p M$ depends on local mole fraction, which is not correct. This is the reason molar-based expression is chosen in this paper.

The above derivations show that under the assumptions of constant pressure and ideal gas behavior, heat flux is independent of the concentration gradients.

Equations (11), (12) and (17) are identical to that in standard textbook for Stefan problem [32], except now the boundary values for the gaseous phase at the interface are $T_1$, $n_{v1}$, $n_{a1}$, rather than determined by the liquid surface temperature $T_s$. In addition to interface conditions Eqs. (7-9), the energy balance at the interface leads to

$$h_w(T_w - T_s) - M_v \dot{N}\Gamma(T_s) = q = C_p T\dot{N} - k\frac{dT}{dz} \tag{18}$$

The boundary conditions in the gas phase away from the interface are specified at a distance $d_v$ away from the interface (see Fig.1)

$$z = d_w + d_v \quad T = T_o, \ P = P_o, n_v = n_{vo} \tag{19}$$

Since Eq.(18) is identical to that of a pure single phase, the previously obtained solution is still valid [42],

$$T(z) = \frac{T_o[exp((z-d_w)/d_c)-1]+T_1[exp(d_v/d_c)-exp((z-d_w)/d_c)]}{exp(d_v/d_c)-1} \quad (20)$$

where $d_c = k/(C_p \dot{N})$ is a characteristic length. Note that the Maxwellian distributions in Eq.(4) imply ideal gas for both species, $c_p \dot{m} = \frac{5}{2} R \dot{N}$ is a constant and hence $d_c$ can be treated as a constant.

Substituting the above temperature distribution into Eq.(18) yields the heat flux as

$$q = C_p T_1 \dot{N} - C_p \dot{N} \frac{T_o - T_1}{exp(d_v/d_c)-1} \quad (21)$$

Since the net mass flux of the non-condensable gas is zero, Eq.(12) leads to,

$$u = -\frac{n}{n_a} D \frac{d\chi_v}{dz} \quad (22)$$

Substituting the above expression to Eq.(11) yields

$$\dot{N} = \dot{N}_v = -\frac{n}{1-\chi_v} D \frac{d\chi_v}{dz} \quad (23)$$

which is identical to the Stefan problem [32]. The above relation can also be readily derived from Eqs. (11) and (14), which shows inherent consistency with the continuity requirement. In literature, mass transfer modeling often assumes that the flow is incompressible, which implies that velocity is a constant according to Eq.(14). However, ideal gas law implies that the molar density is a function of temperature when pressure is constant, which varies with z according to Eq.(20). Consequently, u is not a constant. In fact, evaporation is not just driven by temperature gradient, but also by the density gradient. Thus, constant density assumption is in fundamental conflict with mass diffusion, and it is entirely unnecessary.

We further write Eq.(23) as

$$\frac{\dot{N}RT}{DP} = -\frac{1}{1-\chi_v} \frac{d\chi_v}{dz} \quad (24)$$

With the temperature distribution as given by Eq.(20), the solution for the distribution of the vapor mole fraction $\chi_v(z)$ is

$$\ln\left(\frac{1-\chi_v(z)}{1-\chi_{v1}}\right) = \ln\left(\frac{\chi_a(z)}{\chi_{a1}}\right) =$$

$$\frac{\dot{N}R}{DP}\frac{T_o\{d_c[exp((z-d_w)/d_c)-1]-(z-d_w)\}+T_1\{exp(d_v/d_c)(z-d_w)-d_c[exp((z-d_w)/d_c)-1]\}}{exp(d_v/d_c)-1} \qquad (25)$$

where $\chi_{v1} = \frac{n_{v1}}{n_{v1}+n_{a1}}$. Note that both $n_{v1}$ and $n_{a1}$ are to be determined based on the boundary conditions. Setting $z=d_w+d_v$, Eq.(25) connects the mole fraction at the interface to that at the outer boundary as

$$ln(1+B) = \frac{\dot{N}R}{DP}\frac{T_o[d_c exp(d_v/d_c)-d_c-d_v]+T_{v1}[(d_v-d_c)exp(d_v/d_c)+d_c]}{exp(d_v/d_c)-1} \qquad (26)$$

where $B = \frac{\chi_{v1}-\chi_{vo}}{1-\chi_{v1}}$ is the blowing parameter [32] based on molar fraction rather than the conventional mass fraction.

## 3. Results and Discussion

The above set of equations are highly nonlinear and difficult to solve. This difficult is overcome by reducing unknowns into $T_s$ and $\dot{N}$, and expressing all other quantities using these two unknowns. The final two equations solved are one of Eq.(18), together with $\chi_v + \chi_a = 1$. The mole fraction of the non-condensable gas at the boundary $\chi_{ao}$ is specified as follows. First, for the given temperature $T_o$, the corresponding saturation density of the pure water vapor is calculated using the following empirical relation [56]

$$\rho_{os} = 5.018 + 0.32321 \times t_o + 8.1847 \times 10^{-3}t_o^2 + 3.1243 \times 10^{-4}t_o^3 \qquad (27)$$

where $t_o=T_o-273$ (°C) and $\rho_s$ is in [g]. This fit works well for $t_s$ in [0, 40] °C. The ideal gas law is then used to calculate the corresponding saturation vapor pressure at the boundary $P_{os}$, to be consistent with the displaced Maxwellian distribution, despite water vapor does not exactly obey the ideal gas law.

For the case of a small amount of non-condensable gas, the actual pressure at the boundary $P_o$ is then set slightly above the saturation vapor pressure, with the difference between $P_o$ and $P_{os}$ representing the air partial pressure at the boundary. From this, the air and water vapor mole fractions can be calculated.

For the case of evaporation and condensation into ambient air, the total ambient pressure $P_o$ is set. From $P_o$ and $P_{os}$, one can calculate the saturation water vapor mole fraction at the boundary. The actual water vapor mole fraction $\chi_{vo}$ is then determined by the relative humidity.

The numerical examples shown below uses the following parameters: $C_p=2.5R$, $\alpha_v=\alpha_a=0.5$, $D=2.5\times10^{-5}$ m²/s, $k=0.026$ W/m-K, $M_v=0.0018$ kg/mole, $\Gamma =2.45$ MJ/kg, and molar mass of non-condensable gas $M_a=0.0029$ kg/mole. These values are representative of water in air, although the specific heat of ideal gas is used rather than actual water vapor specific heat. This choice is because the author had identified before that an interfacial cooling effect exists because of the mismatch of convective heat flux in the bulk region with the heat flux leaving the interface due to the difference in the molecules' angular distributions [42].

### 3.1 Evaporation and Condensation in the Presence of Non-condensable Gas

Figures 2(a-c) illustrate the effect of small amount of non-condensable gas on evaporation, for the conditions as given in the figure. At the water-vapor interface, a temperature discontinuity exists, the temperature distribution in the vapor phase is inverted, and the vapor temperature is lower than that of the liquid surface, for both pure water vapor or in the presence of the non-condensable gas. The water vapor density also shows a discontinuity at the interface [Fig.2(b)]. In a previous work [42], the author discussed the cause of the temperature and density discontinuities, and the refrigeration effect at the interface. The effect of the non-condensable gas is small since most of gas is swept away from the interface, as the air concentration shows in Fig.2(c).

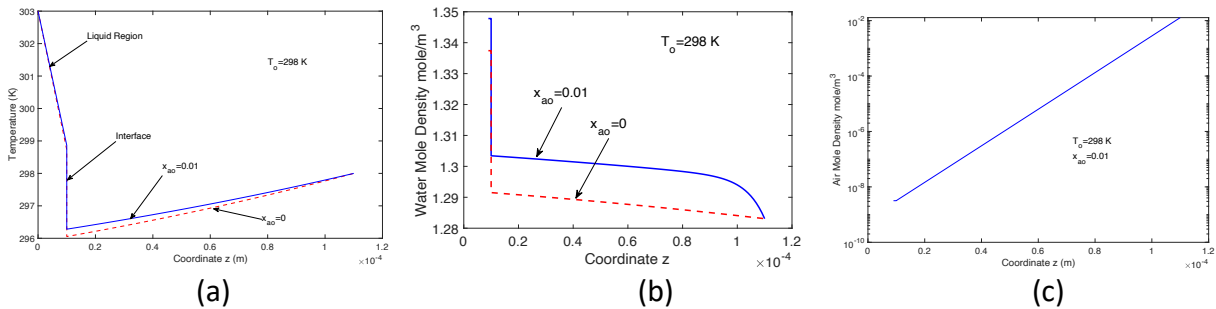

(a) (b) (c)

**Figure 2 Evaporation in the presence of non-condensable gas.** Distributions of (a) temperature, (b) water vapor density (values at liquid surface are artificially plotted into the liquid region for clarity), and (c) non-condensable gas density. $P_o$=3210.7 Pa, $d_w$=10 μm, and $d_v$=100 μm.

However, the situation drastically changed for condensation in the presence of non-condensable gas, as shown in Fig.3(a-c). Although a temperature inversion and discontinuity also exist at the interface for phase water ($x_{a0}$=0), the trend of the vapor phase temperature distribution completely changes despite the fact that the outer boundary has the same amount of air. This is attributed to the accumulation of the non-condensable gas near the interface, as shown in Fig.3(c).

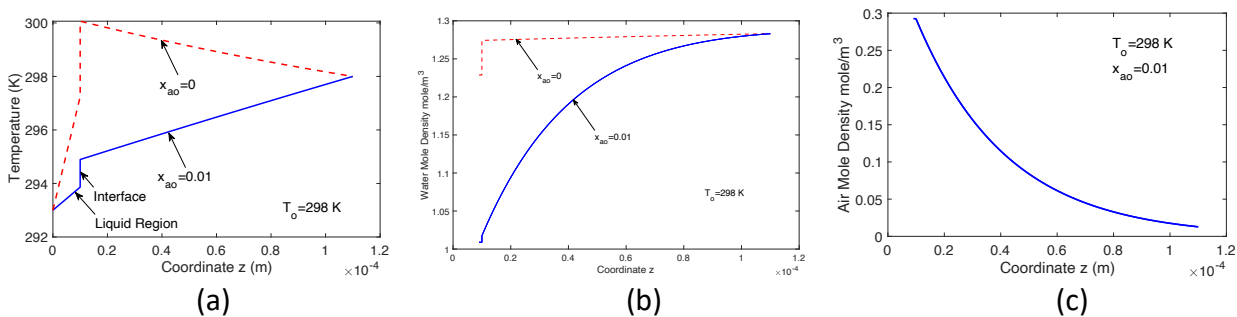

(a) (b) (c)

**Figure 3 Condensation in the presence of non-condensable gas.** Distributions of (a) temperature, (b) water vapor density, and (c) non-condensable gas density. $P_o$=3179 Pa, $d_w$=10 μm, and $d_v$=100 μm.

Figure 4(a) plots the condensation and evaporation rates as a function of the mole fraction of the non-condensable gas at the outer boundary, which further shows the different impacts of non-condensable gas on condensable and evaporation. While evaporation rate is little impacted, the condensation rate is significantly impacted by the presence of minute amount of non-condensable gas. Even when the mole fraction of non-condensable gas at the outer boundary is at $x_{oa}=10^{-5}$, the condensation rate is still only 60% of the condensation rate of pure water, because, the air mole fraction at the liquid-vapor interface still reaches a high value, as shown in Fig.4(b). We should mention that the modeling here does not specify the amount of non-condensable gas in the region. The mole-fraction non-condensable gas at the outer boundary $x_{ao}$ determines the amount of non-condensable in the region.

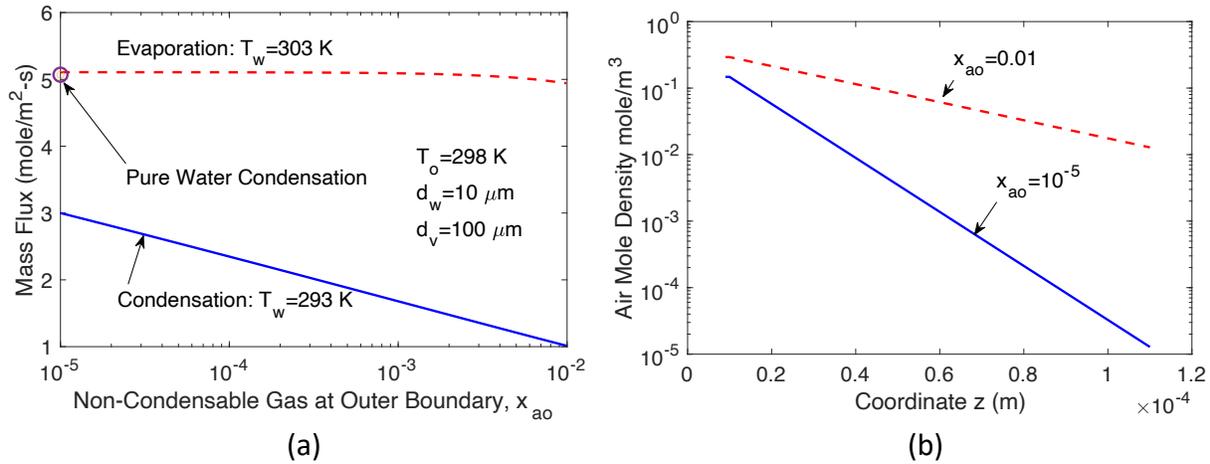

(a)          (b)

Figure 4. (a) Comparison of condensation and evaporation mass flux as a function of the mole fraction of non-condensable gas at outer boundary, and (b) distribution of non-condensable gas for two different outer boundary conditions. Even though the outer boundary values differ by three-orders of magnitude, the difference in the non-condensable gas concentration at the liquid-vapor interface is much smaller.

### 3.2 Evaporation into and Condensation from Air
Figures 5(a-c) show the distributions of temperature, water vapor and air density for evaporation into and condensation from air. Due to the presence of large amounts of air, which contributes to heat transfer despite its zero net mass flux, the temperature distributions are no longer inverted for evaporation or condensation. There are still differences between evaporation and condensation, however. For evaporation, a small interfacial temperature discontinuity exists, while for condensation, no discontinuity was discernible. Naturally, more air accumulates near the interface for condensation, while for evaporation, the density of non-condensable gas at the interface is less than at the outer boundary. The evaporation mass flux is $\dot{N} = 0.11 \text{ mole}/(\text{m}^2\text{s})$, 28% larger than the condensation flux of $\dot{N} = 0.086 \text{ mole}/(\text{m}^2\text{s})$, due to the slight asymmetry

caused by the opposite directions of the average velocity of the gas, as reflected in the air density distribution in Fig.5(c).

In the above examples, the boundary conditions are set such that both temperature and concentration differences at boundaries exist. Fig.6 shows evaporation driven by humidity differences only, with the two boundaries set at same temperature. In this case, transport is completely due to the concentration gradient. There is clearly an inverted temperature profile, as the interface is colder than both the liquid and the air. Heat also conducts back from the air side to the interface. There is also a small temperature discontinuity at the interface.

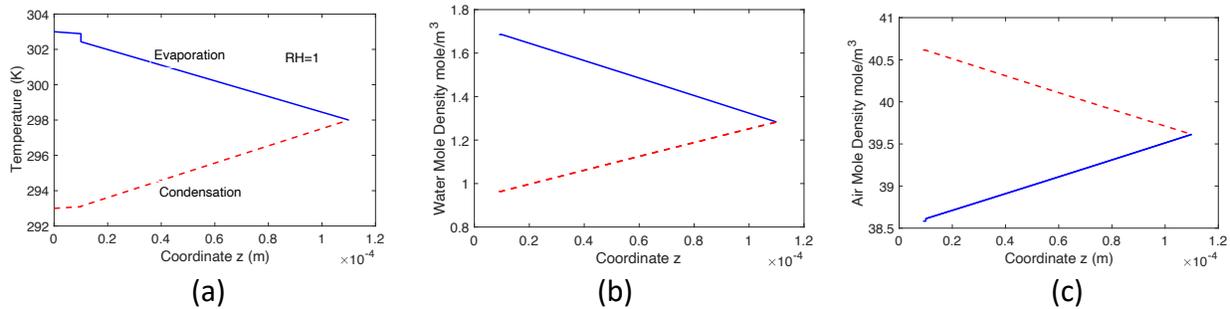

(a)  (b)  (c)

**Figure 5 Evaporation into and condensation from saturated air.** Distributions of (a) temperature, (b) water vapor density, and (c) air density. $P_o$=1 atm, $d_w$=10 μm, and $d_v$=100 μm.

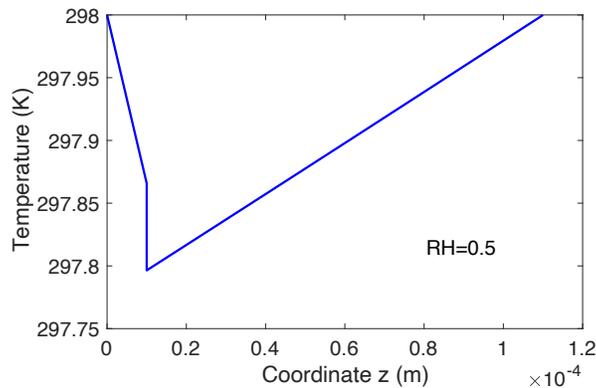

Figure 6. Evaporation driven by humidity difference only, with the wall and the ambient fixed at same temperature.

## 4. Conclusions

Although it is known that the presence of non-condensable gases has significant impacts on the phase change heat transfer processes, especially condensation, it was not previously known if the presence of non-condensable gas will impact interfacial temperature and density discontinuities and temperature inversion, predicted in the past based on the kinetic theory. Here, the author shows that recently derived interfacial mass and heat flux conditions for phase-change heat transfer can be adapted to include the presence of non-condensable gas. These boundary conditions enable one to use the continuum descriptions in both the liquid and the vapor phase, and to calculate the discontinuities of temperature and densities at the interface.

Examples are given for one-dimensional evaporation and condensation in the presence of non-condensable gas, showing the difference between evaporation and condensation in the presence of a small amount of non-condensable gas and for evaporation into and from air, i.e., large amount of non-condensable gas. The presence of a minute amount of non-condensable gas does not have a strong effect on evaporation. Temperature and density discontinuities exist at the interface, and temperature inversion happens, similar to that of evaporation of pure vapor. The situation changes considerably for condensation, for which the presence of a small amount of non-condensable gas flips the temperature distribution, i.e., no inversion happens. For evaporation into and condensation from air, no temperature inversion happens due to the presence of large amount of air. However, for purely humidity driven evaporation, temperature inversion does occur. These new boundary conditions provide a foundation for better modeling of phase change process in the presence of non-condensable gases, which is a big concern in applications such as heat pipes and power plant condensers, and is inevitable for applications such humidification and dehumidification, air-gap distillation and solar-interfacial evaporation, atmospheric water harvesting, cloud formation, etc.

**Acknowledgments:**
This work is dedicated to the special volume in honor of Professor John H. Lienhard, IV. His books with the author's colleague Professor John H. Lienhard, V and with the author's PhD advisor Chang-Lin Tien [Refs.25 and 32] taught me a lot on mass transfer and kinetic theory. The author would like to thank Professor John H. Lienhard V, Adela Li, Simo Pajovic, Xuanjie Wang, and James H. Zhang for proofreading and commenting on the manuscript, and funding from MIT J-WAFS, UM6P, and MIT Bose Award.

**Nomenclature**

| | |
|---|---|
| f | distribution function |
| $\dot{m}$ | mass flux [g/m$^2$-K] |
| M | Molar mass [g/mole] |
| n | molecular molar density [mole/m$^3$] |
| $\dot{N}$ | number flux, [1/m$^2$-s] |
| q | heat flux [W/m$^2$] |
| R | Universal gas constant [J/mole-K] |
| T | Temperature [K] |
| u | average velocity [m/s] |
| v | random velocity [m/s] |
| z | coordinate direction |

**Greek**

| | |
|---|---|
| $\alpha$ | accommodation coefficient |
| $\rho$ | density [g/m$^3$] |
| $\Gamma$ | latent heat [J/kg] |
| $\tau$ | relaxation time [s] |

**Superscript**

| | |
|---|---|
| . | rate |

**Subscript**

| | |
|---|---|
| 1 | vapor phase near the liquid-vapor interface |
| a | Noncondensable phase |
| i | i-th species |
| s | saturated interface |
| v | vapor phase |
| z | z-direction |

**References**

[1] W. Nusselt, Die Oberflachenkondensation des Wasserdampfes, Z. Ver. Dt. Ing. **60**, 569 (1916).
[2] V. P. Carey, *Liquid-Vapor Phase-Change Phenomena: An Introduction to the Thermophysics of Vaporization and Condensation Processes in Heat Transfer Equipment*, 3rd ed. (CRC Press, Boca Raton, 2020).
[3] J. H. Lienhard, Things we don't know about boiling heat transfer: 1988, International Communication on Heat Transfer **15**, 401 (1988).
[4] V. K. Dhir, Mechanistic prediction of nucleate boiling heat transfer-achievable or a hopeless task?, J Heat Transfer **128**, 1 (2006).
[5] H. J. Cho, D. J. Preston, Y. Zhu, and E. N. Wang, Nanoengineered materials for liquid-vapour phase-change heat transfer, Nat Rev Mater **2**, 16092 (2016).

[6]  A. H. Persad and C. A. Ward, Expressions for the Evaporation and Condensation Coefficients in the Hertz-Knudsen Relation, Chem Rev **116**, 7727 (2016).

[7]  W. J. Minkowycz and E. M. Sparrow, Condensation heat transfer in the presence of noncondensables, interfacial resistance, superheating, variable properties, and diffusion, Int J Heat Mass Transf **9**, 1125 (1966).

[8]  J. Huang, J. Zhang, and L. Wang, Review of vapor condensation heat and mass transfer in the presence of non-condensable gas, Appl Therm Eng **89**, 469 (2015).

[9]  J. De Li, M. Saraireh, and G. Thorpe, Condensation of vapor in the presence of non-condensable gas in condensers, Int J Heat Mass Transf **54**, 4078 (2011).

[10]  P. F. Peterson, V. E. Schrock, and T. Kageyama, Diffusion Layer Theory for Turbulent Wapor Condensation With Noncondensable Gases, J Heat Transfer **115**, 998 (1993).

[11]  W. Fu, X. Li, X. Wu, and M. L. Corradini, Numerical investigation of convective condensation with the presence of non-condensable gases in a vertical tube, Nuclear Engineering and Design **297**, 197 (2016).

[12]  H. Ghasemi, G. Ni, A. M. Marconnet, J. Loomis, S. Yerci, N. Miljkovic, and G. Chen, Solar steam generation by heat localization, Nat Commun **5**, 1 (2014).

[13]  P. Tao, G. Ni, C. Song, W. Shang, J. Wu, J. Zhu, G. Chen, and T. Deng, Solar-driven interfacial evaporation, Nat Energy **3**, 1031 (2018).

[14]  H. Kim, S. Yang, S. R. Rao, S. Narayanan, E. A. Kapustin, H. Furukawa, A. S. Umans, O. M. Yaghi, and E. N. Wang, Water harvesting from air with metal-organic frameworks powered by natural sunlight, Science (1979) **356**, 430 (2017).

[15]  T. Defraeye, Advanced computational modelling for drying processes - A review, Appl Energy **131**, 323 (2014).

[16]  Y.-P. Pao, Evaporation in a vapor-gas mixture, J Chem Phys **59**, 6688 (1973).

[17]  S. Brull, V. Pavan, and J. Schneider, Derivation of a BGK model for mixtures, European Journal of Mechanics, B/Fluids **33**, 74 (2012).

[18]  E. L. Walker and B. S. Tanenbaum, Investigation of kinetic models for gas mixtures, Physics of Fluids **11**, 1951 (1968).

[19]  P. Andries, K. Aoki, and B. Perthame, A Consistent BGK-Type Model for Gas Mixtures, J Stat Phys **106**, (2002).

[20]  M. Groppi, S. Monica, and G. Spiga, A kinetic ellipsoidal BGK model for a binary gas mixture, EPL **96**, (2011).

[21]  M. Pirner, A review on BGK models for gas mixtures of mono and polyatomic molecules, Fluids **6**, (2021).

[22]  S. Takata and K. Aoki, Two-surface problems of a multicomponent mixture of vapors and noncondensable gases in the continuum limit in the light of kinetic theory, Physics of Fluids **11**, 2743 (1999).

[23]  T. Soga, Kinetic analysis of evaporation and condensation in a vapor-gas mixture, Physics of Fluids **25**, 1978 (1982).

[24]  C. Cercignani, *Rarefied Gas Dynamics: From Basic Concepts to Actual Calculations* (Cambridge University Press, 2000).

[25]  C. L. Tien and I. J. H. Lienhard, *Statistical Thermodynamics* (Holt, Renehard, and Winston Inc., New York City, 1971).

[26] M. Li, C. Huber, Y. Mu, and W. Tao, Lattice Boltzmann simulation of condensation in the presence of noncondensable gas, Int J Heat Mass Transf **109**, 1004 (2017).

[27] Z. Liang and P. Keblinski, Molecular simulation of steady-state evaporation and condensation in the presence of a non-condensable gas, Journal of Chemical Physics **148**, (2018).

[28] J. Gonzalez, J. Ortega, and Z. Liang, Prediction of thermal conductance at liquid-gas interfaces using molecular dynamics simulations, Int J Heat Mass Transf **126**, 1183 (2018).

[29] J. De Li, CFD simulation of water vapour condensation in the presence of non-condensable gas in vertical cylindrical condensers, Int J Heat Mass Transf **57**, 708 (2013).

[30] W. Wang, B. Li, X. Wang, B. Li, and Y. Shuai, Construction of a Numerical Model for Flow Flash Evaporation with Non-Condensable Gas, Applied Sciences (Switzerland) **13**, (2023).

[31] Z. Lan, R. Wen, A. Wang, and X. Ma, A droplet model in steam condensation with noncondensable gas, International Journal of Thermal Sciences **68**, 1 (2013).

[32] H. Lienhard IV and H. Lienhard V, *A Heat Transfer Textbook*, 6th ed. (https://ahtt.mit.edu/, 2024).

[33] G. Fang and C. A. Ward, Temperature measured close to the interface of an evaporating liquid, Physcal Review E **59**, 417 (1999).

[34] C. A. Ward and D. Stanga, Interfacial conditions during evaporation or condensation of water, Physical Review E **64**, 051509 (2001).

[35] V. K. Badam, V. Kumar, F. Durst, and K. Danov, Experimental and theoretical investigations on interfacial temperature jumps during evaporation, Exp Therm Fluid Sci **32**, 276 (2007).

[36] P. Jafari, A. Amritkar, and H. Ghasemi, Temperature discontinuity at an evaporating water interface, Journal of Physical Chemistry C **124**, 1554 (2020).

[37] R. Schrage, *A Theoretical Study of Interphase Mass Transfer* (Columbia University Press, 1953).

[38] W. G. Vincenti and C. H. Kruger, *Physical Gas Dynamics* (Kriger Pub. Co., 1975).

[39] G. Chen, *Nanoscale Energy Transport and Conversion: A Parallel Treatment on Electrons, Molecules, Phonons, and Photons* (Oxford University Press, 2005).

[40] A. H. Persad and C. A. Ward, Expressions for the evaporation and condensation coefficients in the Hertz-Knudsen relation, Chem Rev **116**, 7727 (2016).

[41] G. Chen, On the molecular picture and interfacial temperature discontinuity during evaporation and condensation, Int J Heat Mass Transf **191**, (2022).

[42] G. Chen, Interfacial cooling and heating, temperature discontinuity and inversion in evaporation and condensation, Int J Heat Mass Transf **218**, (2024).

[43] G. Chen, On paradoxical phenomena during evaporation and condensation between two parallel plates, Journal of Chemical Physics **159**, (2023).

[44] G. Chen, *Nanoscale Energy Transport and Conversion: A Parallel Treatment of Electrons, Molecules, Phonons, and Photons: A Parallel Treatment of Electrons, Molecules, Phonons, and Photons* (Oxford University Press, 2005).

[45] G. Chen, Diffusion-transmission interface condition for electron and phonon transport, Appl Phys Lett **82**, 991 (2003).

[46] S. Kosuge, Model Boltzmann equation for gas mixtures: Construction and numerical comparison, European Journal of Mechanics, B/Fluids **28**, 170 (2009).

[47] K. Aoki, S. Takata, and S. Kosuge, Vapor flows caused by evaporation and condensation on two parallel plane surfaces: Effect of the presence of a noncondensable gas, Physics of Fluids **10**, 1519 (1998).

[48] S. Boscarino, S. Y. Cho, M. Groppi, and G. Russo, BGK models for inert mixtures: comparison and applications, Kinetic & Related Models **14**, 895 (2021).

[49] V. Garzó, A. Santos, and J. J. Brey, A kinetic model for a multicomponent gas, Physics of Fluids A **1**, 380 (1989).

[50] M. 1990- Pirner, *Kinetic Modelling of Gas Mixtures* (n.d.).

[51] S. S. Sazhin, I. N. Shishkova, A. P. Kryukov, V. Y. Levashov, and M. R. Heikal, Evaporation of droplets into a background gas: Kinetic modelling, Int J Heat Mass Transf **50**, 2675 (2007).

[52] J. Yu and H. Wang, A molecular dynamics investigation on evaporation of thin liquid films, Int J Heat Mass Transf **55**, 1218 (2012).

[53] A. Chandra and P. Keblinski, Investigating the validity of Schrage relationships for water using molecular dynamics simulations, Journal of Chemical Physics **153**, 124505 (2020).

[54] E. Bird, J. Gutierrez Plascencia, P. Keblinski, and Z. Liang, Molecular simulation of steady-state evaporation and condensation of water in air, Int J Heat Mass Transf **184**, 122285 (2022).

[55] G. Vaartstra, Z. Lu, J. H. Lienhard, and E. N. Wang, Revisiting the Schrage Equation for Kinetically Limited Evaporation and Condensation, J Heat Transfer **144**, 080802 (2022).

[56] *Empirical Fit of Saturated Vapor Density versus Celsius Temperature*, (unpublished).